\documentstyle[12pt]{article}

\def\rbf{{\bf r}}
\def\kbf{{\bf k}}
\def\ther{thermodynamic limit }
\def\har{harmonic potential }
\def\om{\omega}
\def\Eq({Eq.~(\ref}
\def\Eqs({Eqs.~(\ref}
\def\al{\alpha}
\def\lab{\label}
\def\'{\prime}
\def\kap{\kappa}

\newcommand{\beq}{\begin{equation}}
\newcommand{\eeq}{\end{equation}}
\newcommand{\beqa}{\begin{eqnarray}}
\newcommand{\eeqa}{\end{eqnarray}}

\begin{document}


\pagestyle{empty}
\begin{center}
{\bf
Bose-Einstein Condensation in a Harmonic Potential
\vspace*{0.25cm}\\}
{W. J. Mullin\\}
{\it Department of Physics and Astronomy, University of
Massachusetts,\\
Amherst, MA 01003, USA\vspace*{0.25cm}\\}

\vskip 24pt
\end{center}
\vskip 12pt
\centerline{\bf Abstract}
\vskip 12pt \baselineskip 16pt We examine several features of 
Bose-Einstein condensation (BEC) in an external harmonic potential 
well.  In the thermodynamic limit, there is a phase transition to a 
spatial Bose-Einstein condensed state for dimension D$\geq 2$.  The 
thermodynamic limit requires maintaining constant average density by 
weakening the potential while increasing the particle number $N$ to 
infinity, while of course in real experiments the potential is fixed 
and $N$ stays finite.  For such finite ideal harmonic systems we show 
that a BEC still occurs, although without a true phase transition, below a 
certain ``pseudo-critical'' temperature, even for D=1.  We study the 
momentum-space condensate fraction and find that it vanishes as \( 
1/\sqrt{N} \) in any number of dimensions in the thermodynamic limit.  
In D$\leq 2$ the lack of a momentum condensation is in accord with the 
Hohenberg theorem, but must be reconciled with the existence of a 
spatial BEC in D$= 2$.  For finite systems we derive the \( N
\)-dependence of the spatial and momentum condensate fractions and the 
transition temperatures, features that may be experimentally testable.  
We show that the \( N \)-dependence of the 2D ideal-gas transition 
temperature for a finite system cannot persist in the interacting case 
because it violates a theorem due to Chester, Penrose, and Onsager.

\vskip 12pt

\pagestyle{myheadings}
\vfill\eject

\baselineskip 20pt

\centerline{\bf I.~~~Introduction}
\vskip 12pt

Bose-Einstein condensation (BEC) has been observed recently in several 
laboratories\( ^{\ref{Corn},\ref{Ketb}} \) using magnetic traps to 
confine and cool alkali atoms.  Related experiments on lithium have 
been reported by third group\( ^{\ref{Bra}} \).  The number of atoms 
involved ranged from a few thousand to a few million, in potential 
wells that were to a good approximation anisotropic harmonic 
oscillator potentials.  Such a spatial BEC gives rise to several 
theoretical questions.

First, did these recent experiments observe a true phase transition to 
the BEC state?  The answer is, obviously, no, because a true phase 
transition, with nonanalytic thermodynamic functions, requires taking 
the number of particles and the volume to infinity while keeping the 
density constant.  Of course, no real system ever has such properties, 
but in most homogeneous macroscopic systems the thermodynamic limit is 
a good approximation to the experimental situation in which boundaries 
seem to play a relatively unimportant role.  In the magnetic traps, 
not only is the number of particles quite small, compared to the usual 
case, but the ``boundary,'' formed by a quadratic potential well, 
extends literally throughout the whole system.  In order to take the 
thermodynamic limit in such a system it is necessary to weaken the 
potential so that, as the number of particles increases, the average 
density remains constant.  This is well-defined mathematically, but is 
of course physically unrealizable.  On the other hand, taking the box 
size to infinity in the homogeneous case is also unrealized 
experimentally.  One can argue that the situation there is {\em not} 
qualitatively different from a gas in a harmonic trap.  For the ideal 
gas all that matters is the density of states, and the thermodynamic 
limit simply carries that to a continuum in each case.

The above discussion leads one to ask if the experiments then observed 
a real Bose-Einstein condensation.  The answer seems surely 
positive that they did see a macroscopic number of particles 
occupying the lowest harmonic oscillator state.  Moreover the 
transition occurs quite abruptly as temperature is lowered.  This 
result is in accordance with the findings of several authors\( 
^{\ref{Osb}-\ref{Gro}} \) that in finite homogeneous Bose systems 
there is an accumulation point or ``pseudo-critical'' temperature 
where the increase in the chemical potential slows and the number of 
particles in the ground state begins increasing rapidly.  Many 
different definitions of this accumulation point have been offered for 
an ideal gas -- all of which approach the true phase transition point 
in the thermodynamic limit.

Thirdly, can a pseudo-transition occur in cases where there is no real 
phase transition in the thermodynamic limit?  It can.  One can show\( 
^{\ref{Mul}} \) that there is a real transition in the 
harmonic potential for an ideal gas in any dimension greater than or 
equal to two.  However, although there is no real transition in one 
dimension (1D), there is a pseudo-transition\( ^{\ref{Mul},\ref{Ket}} 
\) that occurs at temperature that would go to zero as
\( 1/\ln N \) in the thermodynamic limit.

The results quoted in the last paragraph bring up yet another 
question.  How can there be a transition in 2D in the harmonic 
potential when the well-known Hohenberg theorem\( ^{\ref{Hoh}} \) says 
that there is no BEC transition in that number of dimensions?  What the 
Hohenberg theorem actually says is that there can be no BEC into the 
\( k=0 \) state where \( k \) is wave number.  Despite the restriction 
to \( k\)-states, this theorem would seem to be relevant to the case 
of a BEC into a harmonic oscillator ground state, because there is yet 
another theorem (CPO theorem), due to Chester\(^{\ref{Chesb},
\ref{Chesa}} \), based on a lemma of Penrose and Onsager\( 
^{\ref{Pen}} \), that notes that there can be no BEC into {\em any} 
single-particle state unless one occurs into a \( k=0 \) state.  So it 
seems that the 2D ideal gas transition ought not to be allowed!  This 
situation has arisen before and the answer found\( 
^{\ref{Reh},\ref{Mas}} \): The CPO theorem requires that the density 
be finite everywhere.  In the thermodynamic limit, the density of an 
ideal gas becomes infinite at the origin in the harmonic oscillator 
problem, which negates the validity of the CPO theorem.  So there can 
be and is a BEC into the harmonic oscillator ground state in 2D in the 
thermodynamic limit.

But the Hohenberg theorem does {\em not} depend on the finiteness of 
the density for {\em its} validity.  So it must still be valid to say 
that, in the thermodynamic limit, there is {\em no} BEC into the \( 
k=0 \) state for the 2D oscillator problem, while there {\em is} one 
for the lowest oscillator state; that is, there is a spatial 
condensation but not a momentum condensation.  We see then that we 
must define two condensate numbers: \( n_{0} \), the number of 
particles in the lowest oscillator state, and \( f_{0} \), the number 
of particles in the zero-momentum state.  If there are \( N
\) particles in the system, then for the 2D ideal gas at non-zero 
temperature in the thermodynamic limit, \( f_{0}/N \) approaches zero, 
while \( n_0/N \) remains a finite fraction.  Basically these two 
quantities are Fourier transforms of one another, and, despite the 
presence of the harmonic potential, it is still possible to have 
particles in the lowest \( k\) state.  This peculiar relationship between 
them has not been noted previously.

Our calculations are for the ideal gas.  In the thermodynamic limit 
the density becomes infinite at the origin in a harmonic potential.  
If a hard-core interparticle repulsion were included, the density 
would not become infinite, the CPO theorem would apply and the 2D BEC 
in a harmonic potential would disappear in the thermodynamic limit.  
It is easy to extend the ideal gas calculations to include a 
mean-field interaction$^{\ref{Klepa},\ref{Klepb}}$, but to the 
author's knowledge no one has yet done the 2D calculation for a system 
with hard-core interactions in a harmonic potential.\( ^{\dag} \)
\footnotetext{ \( ^{\dag} \)A 3D calculation of the transition 
temperature for a gas in a harmonic potential with hard core 
interactions has been done recently by path-integral Monte Carlo 
methods\( ^{\ref{Krauth}} \).} Because the actual experiments are done 
with just a few particles, the ideal gas should become a fair 
approximation to the real experiments, and the density divergence of 
the ideal gas should never become an issue.  Thus our calculations of 
$n_0$, \( f_{0} \), and the transition temperature might be relevant 
to experiment.\( ^{\ddag} \)\footnotetext{\( ^{\ddag} \)As pointed out 
recently in Ref.~\ref{Ket}, experiments on systems with reduced 
dimensionlity (1D or 2D) are entirely possible in magnetic traps.} 
However, the CPO theorem tells us that the pseudo-transition 
temperature would vanish were we to take the thermodynamic limit.  Thus 
the \( N \)-dependence of the pseudo-transition temperature must be 
{\em different} from that of the ideal gas, meaning it must depend 
more weakly on particle number than for the ideal gas.  This is in 
contrast to the 3D case where the ideal gas transition temperature 
seems to describe experiment quite well\( ^{\ref{Tcexper}} \). A full 
hard-core calculation of the transition temperature in 2D would of 
course be useful.

In Sec.~II, we review the treatment of the Bose condensation in the 
thermodynamic limit.  We look at the finite case in more detail in 
Sec.~III. Sec.~IV examines the momentum distribution and its relation 
to the distribution in the harmonic states.  Further discussion 
occurs in Sec.~V. An Appendix gives some mathematical details and 
makes connection with previous work.

\vskip 12pt

\centerline{\bf II.~~Harmonic Bose Systems in the Thermodynamic Limit}
\vskip 12pt

Bose-Einstein condensation in an external \har has been considered 
previously\( ^{\ref{Mul}{\rm -}\ref{Ket},\ref{Mas}{\rm 
-}\ref{Krauth}, \ref{Krauth}{\rm -}\ref{Kirst} } \). We re-examine and 
extend the analysis here.  Consider the two-dimensional system of \( N 
\) noninteracting Bosons with the particles contained in an isotropic 
two-dimensional \har given by 
\beq
\label{Pot} 
U(\rbf)=\frac{1}{2}U_{0}\left(\frac{r}{R}\right)^{2}
\eeq
where \( r^{2}=x^{2}+y^{2} \) and \( R \) is a range parameter for the 
potential.  It is simple to generalize to any number of dimensions 
greater than two.  The energy levels are given
\beq
E_{m_{x},m_{y}}=\hbar \om (m_{x}+m_{y}+1)
\eeq
with \( m_{x}, m_{y} = 0,1,2,\ldots \) The angular frequency is 
\beq
\om = \sqrt{U_{0}\over R^{2}m } 
\eeq
where \( m \) is the mass of a particle.  In the grand ensemble at 
temperature \( T \), the total number of particles \( N \) satisfies 
the equation
\beq
\label{Neq}
N= \sum_{m_{x},m_{y}=0}^{\infty}\frac{1}{e^{\beta 
[\hbar\om(m_{x}+m_{y}+1)-\mu]}-1}
\eeq
where \( \mu \) is the chemical potential, \( \beta=1/k_{B}T \), 
and \( k_{B} \) is the Boltzmann constant. 

If one wishes to take the thermodynamic limit, then one must increase 
the ``volume'' of the system while increasing the number of particles 
with the average density kept constant.  The average density is 
proportional to \( \rho \equiv N/ R^{2} \), where \( R \) is the range 
parameter in the \har of \Eq({Pot}).  To take the \ther requires that 
\( R \) increase, that is, that the potential weaken, as N increases.  
Such a situation has been considered 
previously$^{\ref{Reh},\ref{Mas},\ref{Wid}}$, but might seem peculiar 
when considering real experiments in a potential fixed by external 
magnets.  We follow this procedure, because it is the only way to 
reach the true nonanalytic behavior characteristic of phase 
transitions.  One does not feel so uncomfortable with the \ther in the 
case of a homogeneous particles-in-a-box situation because there it 
seems that all that the \ther does is to remove boundary effects, 
which are mostly negligible anyway in the real bulk experiments.  
Making such a qualitative distinction between the two cases seems 
unwarranted, however.

If \( \rho \) is kept constant we require, in 2D, that
\beq
\om \propto \frac{1}{R}\propto \frac{1}{\sqrt{N}}
\eeq
In \( D \) dimensions, \( \rho \propto \frac{N}{R^{D}}\) and \( \om 
\propto \frac{1}{N^{1/D}} \). 

Introduce 
\beq
\label{T0}
T_{0}^{(2)}\equiv \frac{\hbar }{k_B}\sqrt{\frac{ U_{0}}{m}}\sqrt{\rho}
\eeq
and 
\beq
\al=-\beta \mu + \hbar \om.
\eeq
Then 
\beq
\frac{\hbar \om}{k_B T} =\frac{T_{0}^{(2)}}{T\sqrt{N}}
\eeq
and \Eq({Neq}) can be written as 
\beq
N= \sum_{m_{x},m_{y}}\frac{1}{e^{\frac{T_{0}^{(2)}}{T\sqrt{N}} 
(m_{x}+m_{y})+\al}-1}. 
\eeq

To simplify, one can change the sums over \( m_{x} \) and \( m_{y} \) 
to sums over \( p=m_{x}+m_{y} \) and \(l=m_{x} \). Then, since \( l 
\) no longer appears in the summand, one can do that sum to give
\beq
\lab{2Dpeq}
N= \sum_{p=0}^{\infty}\frac{p+1}{e^{\frac{T_{0}^{(2)}}{T\sqrt{N}}p+\al}-1}.
\eeq
One can use this formula to do numerical calculations of \( \al \) 
for given finite \( N \). We will discuss this procedure in the next 
section. Here we want to take the \ther and it is useful to separate 
off the ground state by writing
\beqa
\lab{n0N'}
N &=& n_{0}+\sum_{p=1}^{\infty}\frac{p+1}
{e^{\frac{T_{0}^{(2)}}{T\sqrt{N}}p+\al}-1}\nonumber\\
&=& n_{0}+\sum_{p=0}^{\infty}\frac{p+2}
{e^{\frac{T_{0}^{(2)}}{T\sqrt{N}}p+\al'}-1}.
\eeqa
where 
\beq
\lab{nzero}
n_0=\frac{1}{e^{\al}-1}.
\eeq
In the last form we have taken \( p\rightarrow p+1 \) and \( 
\al'=\al+ \frac{T_{0}^{(2)}}{T\sqrt{N}}\) to reset the sum from \( p=0\). 

When \( N \) gets large, the states become very closely spaced and we 
can replace the sum by an integral (see Appendix) to good 
approximation to become
\beq
N=n_0 +\int_{0}^{\infty}dp\frac{p+2}
{e^{\frac{T_{0}^{(2)}}{T\sqrt{N}}p+\al'}-1}
\eeq
Changing variables to \( u=p T_{0}^{(2)}/\sqrt{N}T \), we find 
\beqa
\lab{2Dint}
N&=&n_0+N\left(\frac{T}{T_{0}^{(2)}}\right)^{2}\int_{0}^{\infty}du
\frac{u}{e^{u+\al'}-1} + 2\sqrt{N}\left(\frac{T}{T_{0}^{(2)}}\right)
\int_{0}^{\infty}du\frac{1}{e^{u+\al'}-1}\nonumber\\
&=&n_0+N\left(\frac{T}{T_{0}^{(2)}}\right)^{2}F_{2}(\al')
+2\sqrt{N}\left(\frac{T}{T_{0}^{(2)}}\right)F_{1}(\al')
\eeqa
where
\beq
F_{\sigma}(\al)\equiv \frac{1}{\Gamma(\sigma)}\int_{0}^{\infty}du
\frac{u^{\sigma -1}}{e^{u+\al}-1}=\sum_{p=1}^{\infty}\frac{e^{-\al 
p}}{p^{\sigma}}
\eeq
are the Bose integrals\( ^{\ref{Lon}} \); \( \Gamma(\sigma)\) is the 
Gamma function.  The term in \( F_{2} \) is of order \( N \) while the 
term in \( F_{1} \) is of order \( \sqrt{N} \) and can be neglected in 
the thermodynamic limit.  We will however use it in the next section 
to estimate the error made by the \ther form when \( N \) is finite.  
Since \( F_{2} \) has a finite limit for small \( \al \), there is a 
BEC for temperatures lower than some critical temperature.  \( 
F_{2}(\al)\) behaves for small \( \al \) as\( ^{\ref{Lon}} \)
\beq
F_{2}(\al)=\zeta(2)-\al(1-\ln \al)+\ldots
\eeq
where \( \zeta(\sigma)=\sum_{p=1}^{\infty}1/p^{\sigma} \) is 
the Riemann \( \zeta \)-function.  We have \( \zeta(2) 
=\pi^{2}/6 \).  The 2D Bose condensation temperatures \( T_{c}^{(2)} \) 
is given by the equation
\beq
N=N\frac{T_{c}}{T_{0}^{(2)}}F_{2}(0).
\eeq
or 
\beq
\label{Tc}
T_{c}^{(2)}=T_{0}^{(2)}\sqrt{\frac{6}{\pi^{2}}}\approx 
0.78 T_{0}^{(2)}.
\eeq

The occupation of the lowest oscillator state is, for \( T\leq 
T_{c}^{(2)} \),
\beq
\lab{n02D}
\frac{n_0}{N}=1-\left(\frac{T}{T_{c}^{(2)}}\right)^{2}\hspace{2cm} 
{\rm 2D}
\eeq
and zero above \( T_{c}^{(2)} \).  Unlike the 2D homogeneous system 
the 2D oscillator has a BEC phase transition in the thermodynamic limit.

Recently the authors of Ref.~\ref{Kirst} have claimed that the 
harmonic potential is fundamentally different from a homogeneous 
system and that a true phase transition cannot occur in the harmonic 
potential.  We see that this is not correct, although the physical 
conditions for its occurrence (weakening the potential) may seem a bit 
strange.  However, the same arguments could be applied to the system 
of particles in a box.  Real experiments do not occur in an infinitely 
large box and experimentalists do not observe actual phase transitions 
with truly discontinuous functions.  Further, all the particles know 
is the energy levels and their spacing; in each case the limiting 
process is just changing the spacing.  We feel that the two cases are 
not fundamentally different.

In 3D we have 
\beq
\lab{N3D}
N=\sum_{m_{x},m_{y},m_{z}=0}^{\infty}\frac{1}{e^{\frac{T_{0}^{(3)}}
{TN^{1/3}}(m_{x}+m_{y}+m_{z})+\al}-1}
\eeq
where \( T_{0}^{(3)} \) is given by 
\beq
\lab{T03}
T_{0}^{(3)}\equiv \frac{\hbar}{k_B}\sqrt{\frac{ U_{0}}{m}}\rho^{1/3}
\eeq
with \(\rho\equiv N/R^{3}\)
and
\beq
\al=-\beta \mu + 3T_{0}^{(3)}/(2TN^{1/3}).
\eeq
One can again reduce the sum of \Eq({N3D}) to one variable over \( 
p=m_{x}+m_{y}+m_{z} \) to give to 
\beq
\label{Nseries}
N=n_0 + \sum_{p=0}^{\infty}\frac{\frac{1}{2}p^{2}+
\frac{5}{2}p+3}{e^{\frac{T_{0}^{(3)}}{TN^{1/3}}p+\al'}-1}
\eeq
with $\al'=\al+T_0/(TN^{1/3})$.
We have again separated off the \( n_{0} \) term and have reset the 
sum from \( p=0 \).  Changing the sum to an integral (see Appendix) 
gives
\beq
\label{3Danaly}
N=n_0 + N\left(\frac{T}{T_{0}^{(3)}}\right)^{3}F_{3}(\al')+
\frac{5}{2}N^{2/3}\left(\frac{T}{T_{0}^{(3)}}\right)^{2}F_{2}(\al')
+3 N^{1/3}\left(\frac{T}{T_{0}^{(3)}}\right)F_{1}(\al')
\eeq
In the \ther the terms in \( F_{2} \) and \( F_{1} \) are negligible, 
as is the difference between \( \al' \) and \( \al \), 
and the condensate fraction is given by
\beq
\frac{n_0}{N}= 1-\left(\frac{T}{T_{c}^{(3)}}\right)^{3} 
\hspace{2cm} {\rm 3D}
\eeq
where \( T_{c}^{(3)}=T_{0}^{(3)}/(\zeta(3))^{1/3} \).

\Eq({3Danaly}) is very similar to one given recently in 
Ref.~\ref{Grob}, but differs in the coefficients of the various terms 
and in the argument being \( \al' \) rather than \( \al \).  We show 
in the Appendix that the two equations are the same.  Our equation 
also has the same form as that appearing in Ref.~\ref{Haug}.

It is similarly easy to show that the D-dimensional harmonic 
oscillator system has a BEC, for D\( \geq 2 \), at temperature 
\beq
T_{c}^{({\rm D})}=T_{0}^{({\rm D})}\zeta(D)^{-1/D}
\eeq
with \( T_{0}^{({\rm D})} \) given by
\beq
\lab{T0D}
T_{0}^{({\rm D})}\equiv \frac{\hbar}{k_B}\sqrt{\frac{ U_{0}}{m}}
\rho^{1/D}
\eeq
and \( \rho\equiv N/R^{D} \). In D-dimensions (D\( \geq 2 \)) 
the condensate fraction is given by
\beq
\lab{Ddim}
\frac{n_0}{N}=1-\left(\frac{T}{T_{c}^{({\rm 
D})}}\right)^{D}\hspace{2cm} {\rm D\; dimensions}.
\eeq
As D\( \rightarrow \infty\), \(T_{c}^{({\rm D})}\rightarrow 
T_{0}^{({\rm D})} \) and \( n_0/N \) becomes a step function in 
\( T \).

The 1D system is a special case for which there is no condensation in 
the thermodynamic limit. The same procedure as used above leads, in 
the continuum limit, to 
\beq
\lab{1DNeq}
N=N\left(\frac{T}{T_{0}^{(1)}}\right)\int_{0}^{\infty}\frac{1}{e^{u+\al}-1}
=N\left(\frac{T}{T_{0}^{(1)}}\right)F_{1}(\al).
\eeq
with \( T_{0}^{({1})} \) given by \Eq({T0D}) with D\(=1\).  \( 
F_{1}(\al) \) does not approach a finite limit as \( \al\rightarrow 0 
\), but is given exactly by
\beq
F_{1}(\al) = -\ln(1-e^{-\al})
\eeq
which approaches \( -\ln \al \) as \( \al\rightarrow 0 \).  However, 
Refs.~\ref{Mul} and \ref{Ket} have pointed out that, in the finite 
system, the 1D system {\em does} have a BEC below a certain 
pseudo-critical temperature.  We will discuss this case below.

\vskip 12pt

\centerline{\bf III.~~Finite Systems}
\vskip 12pt
To consider finite harmonic systems\( 
^{\ref{Mul}{\rm -}\ref{Ket},\ref{Krauth},\ref{Haug}{\rm -}\ref{Kirst}} 
\) most easily in, say, 2D, we can just compute \( \al \) via 
\Eq({2Dpeq}) by iteration and then compute the number of particles in 
the lowest state from \Eq({nzero}).  Fig.~1 illustrates the 2D 
situation.  The dotted and dashed lines are the exact results for \( 
N=10 \) to \( 10^{4} \) with the \ther (\Eq({n02D})) shown as the 
solid line.

One notes that the exact result is smaller than the value given by 
the infinite \( N \) limit. A better approximation than \Eq({n02D}) 
is the form of \Eq({2Dint}); the \( F_{1} \) term in \Eq({2Dint}) 
aids in giving an estimate of the difference due to finite \( N \). In 
3D a better analytic approximation is given by 
\Eq({3Danaly})\( ^{\ref{Grob},\ref{Haug}{\rm -}\ref{Kirst}} \). 
Since, for small temperature, \( \al =O(1/N) \), the \( F_{\sigma} \) 
behave as 
\beqa
&&F_{1}(\al)\rightarrow -\ln(\al)\nonumber\\
&&F_{2}(\al)\rightarrow \zeta(2) -\al(1-\ln(\al))
\eeqa
and we can show that the condensate fraction is 
roughly
\beq
\frac{n_0}{N}\approx 1-\left(\frac{T}{T_{c}^{(2)}}\right)^{2}
-\frac{T}{T_{0}^{(2)}}\frac{\ln(N)}{\sqrt{N}}
\eeq
The correction is negative so that the value of \( n_0 \) valid for 
finite \( N \) is smaller than the infinite limit value, at least 
when \( \al \) is of order \( 1/N \) as we see in the figure.

For a finite harmonic system it is not particularly appropriate to 
express the pseudo-transition temperature in terms of the density.  
The potential is fixed and scaling it in terms of \( N \) with fixed 
density makes little sense.  We now express it instead in terms of a 
fixed frequency.  The effective or pseudo- transition temperature is 
only a bit less (less by terms of order \( \ln N/\sqrt{N} \)) than the 
infinite-limit transition temperature and can be taken as roughly the 
same.  By \Eqs({T0}) and (\ref{Tc}) the pseudo-transition temperature 
is given by
\beq
\label{Tc2Dfix}
T_{c}^{(2)}=\sqrt{\frac{N}{\zeta(2)}}\frac{\hbar\om}{k_{B}}
\eeq

A similar analysis for three dimensions leads to a pseudo-transition 
temperature a bit less than
\beq
T_{c}^{(3)}=\left(\frac{N}{\zeta(3)}\right)^{1/3}
\frac{\hbar\om}{k_{B}}
\eeq

A pseudo-transition in one-dimension is a special 
case$^{\ref{Mul},\ref{Ket}}$ because there is no real phase 
transition in 1D. Pseudo-transition temperatures can be defined in 
many ways.  Ref.~\ref{Gob} defines six different ways in the 
homogeneous case that all lead to the same $T_c$ for infinite \( N \) 
in cases where there is a real phase transition.  We compare just two 
approaches.  First we follow$^{\ref{Mul}}$ the analysis used in the 
higher dimensional cases: Write
\beq
\lab{1Dsum}
N=\sum_{p=0}^{\infty}\frac{1}{e^{\frac{T_{0}^{(1)}}{TN}p+\al}-1}
\eeq
where \( T_{0}^{(1)} \) is given by 
\beq
T_{0}^{(1)}\equiv \frac{\hbar}{k_B}\sqrt{\frac{ U_{0}}{m}}\frac{N}{R}
\eeq
and \( \al=-\beta\mu +T_{0}^{(1)}/(2TN) \). 
Separate off \( n_0 \) as above to get
\beq
N=n_0+ \sum_{p=0}^{\infty}\frac{1}{e^{\frac{T_{0}^{(1)}}{TN}p+\al'}-1}
\eeq
in which 
\beq
\lab{1Dal}
\al'=\al+\frac{T_{0}^{(1)}}{TN} 
\eeq
and the sum has been reset from \( p=0 \). The continuum limit 
becomes
\beq
\lab{1Dcont}
N=n_0+ N\frac{T}{T_{0}^{(1)}}F_{1}(\al'). 
\eeq
The assumption that \( n_0 \) is of order \( N \) implies, from 
\Eq({nzero}), that \( \al\) is of order \( 1/N \). 
We have, for small \( \al' \),
\beq
\lab{1Dsmall} 
N= n_0 - \frac{T}{T_{0}^{(1)}}\ln(\al')
\eeq
Taking \( \al' \) of order \( 1/N \) and assuming \( N \) large give
\beq
N=n_0+\frac{T}{T_{0}^{(1)}}N\ln N
\eeq
or 
\beq
\lab{1Dn0}
\frac{n_0}{N}=1-\frac{T}{T_{0}^{(1)}/\ln N}
\eeq
This equation has just the form shown in \Eq({Ddim}) with an \( N 
\)-dependent transition temperature. (But, of course, in the case of a 
finite \( N \) and fixed \har {\em all} pseudo-transition 
temperatures are \( N \)-dependent.)  Here we have
\beq
T_{c}^{(1)}=\frac{T_{0}^{(1)}}{\ln N}=\frac{N}{\ln N}\frac{\hbar\om}{k_{B}}
\eeq
in which the first form is appropriate to the case where the density 
\( N/R \) is being kept constant (weakening potential) and shows the 
transition temperature decreasing to zero as the number of particles 
increases so that there is no true phase transition.  However, the 
second form shows that for constant potential as in the real 
experiments the temperature for which a substantial BEC occurs 
actually gets larger as \( N \) increases.

An alternative derivation$^{\ref{Gob},\ref{Ket}}$ comes easily from 
noting that in a BEC one 
can write
\beq
N=n_{0}(\al) +N'(\al)
\eeq
where \( N' \) is the number of particles not in the ground state.  
At 
a real BEC transition temperature \( \al \) is very small, but \( 
n_{0} \) is also still microscopic.  Thus at the transition 
temperature, we have approximately
\beq
N'(0)=N
\eeq
In the case of the 1D system this would imply, from \Eqs({1Dsmall}) 
and (\ref{1Dal}) that \( T_{c}^{(1)}\) is given by
\beq
-\frac{T_{c}^{(1)}}
{T_{0}}\ln\left(\frac{T_{0}^{(1)}}{T_{c}^{(1)}N} \right)=1
\eeq
or to a good approximation for large \( N \),
\beq
T_{c}^{(1)}=\frac{T_{0}^{(1)}}{\ln N}
\eeq
in agreement with the first estimate.

In Fig.~2, we plot the exact quantum result \( n_{0}/N \) from 
\Eq({1Dsum}) as a function of \( T/T_{0}^{(1)} \).  Such a graph is 
appropriate to the case of taking the thermodynamic limit for which 
the density (and therefore \( T_{0}^{(1)} \)) is held constant.  One 
sees that then the transition temperature gets smaller as \( N \) 
increases, indicating that there is no real transition in this system.  
Fig.~3, on the other hand, plots \( n_{0}/N \) as a function of \( 
T\ln N/T_{0}^{(1)} \).  This method of graphing\( ^{\ref{Ket}} \) 
includes all the \( N \) dependence that occurs in \( T_{c}^{(1)} \) 
and is more appropriate for the fixed potential case.  One sees the 
asymptotic approach to an exact straight line for large \( N \) as 
given by \Eq({1Dn0}).

\vskip 12pt

\centerline{\bf IV.~~The Hohenberg and CPO Theorems}
\vskip 12pt The Hohenberg theorem\( ^{\ref{Hoh}} \) states that there 
is no condensation into the \( k=0 \) state in the infinite system in 
two dimensions or less.  The CPO theorem\( ^{\ref{Chesb}{\rm 
-}\ref{Pen}} \) says that there can be no condensation into any single 
particle-state unless there is one into the \( k=0 \) state.  However, 
the CPO theorem relies on the assumption that the density is 
everywhere finite.  This is not the case in a harmonic potential, 
where we will show that the density becomes infinite at the origin in 
the thermodynamic limit.  We first review the Hohenberg theorem as 
presented by Chester\( ^{\ref{Chesb}} \).

Chester has shown how to apply the Hohenberg theorem to a finite, 
inhomogeneous system.  To quote Chester\( ^{\ref{Chesb}}\): `` \ldots 
[F]or any system whatsoever, we can always ask about the mean number 
of particles with momentum zero.  This follows from the general 
principles of quantum mechanics and we are in no way limited, in 
asking this type of question, to systems that are homogeneous.'' We 
consider the following picture.  Assume that the \har extends out to 
radius \( R \), but that there is a hard wall at \( r=R\).  It is 
possible, of course, to find the energy eigenvalues for such a mixed 
potential system, but we can use the eigenvalues of the \har if we 
just assume that \( k_{B}T \) is small enough that particles rarely 
ever actually reach the hard wall.  That is, we assume that \( R \) 
is 
sufficiently large that
\beq
k_{B}T< V(R)
\eeq

The Bogoliubov inequality is 
\beq
\left<\frac{1}{2}[A,A^{\dagger}]\right>\geq 
\frac{k_{B}T|\left<[C,A]\right>|^{2}}{\left<[[C,H],C^{\dagger}]\right>}
\eeq
in which Chester takes 
\beq
A=a_{0}a_{k}^{\dagger}
\eeq
and
\beq
C=\sum_{q}a_{q}^{\dagger}a_{q+k}
\eeq
where \( a_{k}^{\dagger} \) is the creation operator for the state 
with wave number {\bf k}.  To quote Chester once again\( 
^{\ref{Chesb}}\): ``Since we are allowed to use any set of complete 
functions for our second quantization formalism we choose a continuum 
of plane waves.'' Direct calculation of the commutators leads to
\beq
f_{0}f_{k}+\frac{1}{2}(f_{0}+f_{k})\geq 
\frac{(f_{0}-f_{k})^{2}k_{B}Tm}{N(\hbar k)^{2}}
\eeq
where \( f_{k}=<a_{k}^{\dagger}a_{k}> \) is the number of particles in 
the momentum state ${\bf k}$.  This inequality is summed over k values 
from some minimum \( k_{0} \) to maximum \( k_{c} \).  The first term 
on the left can have the sum changed to {\em all} \( k \) values, 
which just increases the inequality and gives \( Nf_{0} \).  The 
second term becomes \( Nf_{0}\Gamma \), where \( \Gamma \) is of order 
unity.  The third term is of order \( N \) and is negligible if \( 
f_{0} \) is assumed to be of order \( N \).  On the right side we can 
drop \(f_{k} \) as negligible compared to \( f_{0} \).  Thus we have
\beq
Nf_{0}\left(1+\frac{\Gamma}{N}\right)\geq 
\frac{f_{0}^{2}mk_{B}T}{N\hbar^{2}}\sum_{k=k_{0}}^{k_{c}}
\frac{1}{k^{2}}
\eeq
And so in 2D, we have (assuming that dropping the \( \Gamma \) term 
does not seriously affect the inequality)   
\beq
N\geq \frac{f_{0}mk_{B}TA}{h^{2}}\ln \frac{k_{c}}{k_{0}}
\eeq
where \( A =\pi R^{2}\) is the area of the system.  We take \( 
k_{0}\propto 1/R \) and \( k_{c}\propto 1/a \), where \( 
a=1/\sqrt{\rho}=\sqrt{R^{2}/N} \).  Then our result for the 2D case is
\beq
\frac{f_{0}}{N}\leq \frac{\hbar^{2}}{ma^{2}k_{B}T}\frac{1}{\ln N}
\eeq

Thus the Hohenberg theorem states that, in the thermodynamic limit, 
the condensate fraction in the \( k=0 \) momentum state must go to 
zero at least as fast as \( 1/\ln N \) for any non-zero temperature.  
We will see that it actually goes to zero faster than this, namely as 
\( 1/\sqrt{N}\).

The CPO theorem goes as follows: Let \( \sigma_{1}(\rbf,\rbf^{\'}) \) 
be the single-particle reduced density matrix in the position 
representation.  Penrose and Onsager's lemma states that if the 
quantity
\beq 
W=\frac{1}{V} \int d\rbf \int d\rbf^{\'}|\sigma_{1}(\rbf,\rbf^{\'})|, 
\eeq 
(where \( V \) is the volume) is not proportional to \( N \), then 
there is no BEC into {\em any} eigenstate of the single-particle 
density matrix.  However, the number of particles in the zero momentum 
state is given by
\beqa
\lab{f0dens}
f_{0}&=&\int d\rbf \int 
d\rbf^{\'}<\kbf =0|\rbf><\rbf|\sigma_{1}|\rbf^{\'}>
<\rbf^{\'}|\kbf =0>\nonumber\\
&=&\frac{1}{V}\int d\rbf \int d\rbf^{\'} \sigma_{1}(\rbf,\rbf^{\'}). 
\eeqa 
Chester proves that a Bose system has a positive-semidefinite 
single-particle density matrix, so that, if \( f_{0}/N \) approaches 
zero as \( N\rightarrow \infty \), then \( W/N \) must also go to 
zero, eliminating the possibility of BEC into any single-particle 
state. 

The weak link in the theorem's proof for our purposes is that the 
Penrose-Onsager lemma requires the average density at each point be 
finite.  That is not the case in the harmonic potential as we will 
show.  Thus the fact that \( f_{0}/N \rightarrow 0\) does {\em not} 
imply that \( n_{0}/N \rightarrow 0\).  In the \ther it is possible, 
for a 2D harmonic potential, to have a condensation into the lowest 
harmonic oscillator state while still satisfying the Hohenberg theorem 
that there be no BEC into the zero momentum state in the thermodynamic 
limit.  We will indeed show that \( f_{0}/N \) approaches zero as \( 
1/\sqrt{N} \).

A direct computation of \( f_{k} \) is straightforward. We have
\beq
\lab{sumsig}
f_{k}=\sum_{m}<{\bf k}|m><m|\sigma_{1}|m><m|{\bf k}>
\eeq
where the \( |m> \) represent the eigenstates of \( \sigma_{1} \), in 
this case the harmonic oscillator eigenstates.  The states \( <{\bf 
k}|m> \) are the Fourier transforms of the harmonic oscillator states.  
These latter functions are, just like the states in configuration 
space, Hermite functions times a Gaussian.  In 1D
\beq
<k|m>= 
\frac{C_{m}\sqrt{2\pi}}{\xi\sqrt{L}}H_{m}(k/\xi)e^{-k^{2}/2\xi^{2}}
\eeq
in which \( L \) is a box length,
\beq
\xi=\sqrt{\frac{m\om}{\hbar}},
\eeq
and 
\beq
C_{m}=\sqrt{\frac{\xi}{\sqrt{2}2^{m}m!}}
\eeq
is the usual wave function normalization factor in configuration 
space. 

The density matrix is given in 2D 
by
\beq
\sigma_{m_{x}m_{y}}\equiv <m_{x}m_{y}|\sigma_{1}|m_{x}m_{y}>= 
\frac{1}{e^{\beta [\hbar \om(m_{x}+m_{y}+1)-\mu]}-1}
\eeq

The result of the indicated calculations is 
\beq
f_{k}= 
\frac{(2\pi)^{2}}{A\xi^{4}}e^{-\left(\kap_{x}^{2}+\kap_{y}^{2}\right)}
\sum_{m_{x},m_{y}=0}^{\infty}
\sigma_{m_{x}m_{y}}C_{m_{x}}^{2}C_{m_{y}}^{2}\left[H_{m_{x}}(\kap_{x})
H_{m_{y}}(\kap_{y})\right]^{2}
\eeq
where \( \kap_{i}\equiv k_{i}/\xi \), and \( A \) is the area. Using 
the 
relations derived in Sec.~II we can rewrite the last equation as
\beqa
\lab{fsubk}
&&f_{k}=\frac{4}{\sqrt{N}k_{B}T_{0}}\frac{\hbar^{2}}{ma^{2}}
e^{-\left(\kap_{x}^{2}+\kap_{y}^{2}\right)}\nonumber\\
&&\hspace{1cm}\times\sum_{m_{x},m_{y}=0}^{\infty}
\sigma_{m_{x}m_{y}}\frac{1}{2^{m_{x}}2^{m_{x}}m_{x}!m_{y}!}
\left[H_{m_{x}}(\kap_{x})H_{m_{y}}(\kap_{y})\right]^{2}
\eeqa
where \( a\equiv \sqrt{1/\rho} \). 

The value of the zero momentum term is
\beq
\lab{fzero}
f_{0}=\frac{4}{\sqrt{N}k_{B}T_{0}}\frac{\hbar^{2}}{ma^{2}}
\sum_{m_{x},m_{y}=0,2,\ldots}^{\infty}
\sigma_{m_{x}m_{y}}\left[\frac{m_{x}!m_{y}!}{2^{m_{x}}2^{m_{x}}
[(m_{x}/2)!(m_{y}/2)!]^{2}}\right]
\eeq

We can find the density at position \( \rbf \) by a similar analysis:
\beq
\rho(\rbf)=<\rbf|\sigma_{1}|\rbf>=e^{-\xi^{2}r^{2}}
\sum_{m_{x},m_{y}=0}^{\infty}
\sigma_{m_{x}m_{y}}C_{m_{x}}^{2}C_{m_{y}}^{2}\left[H_{m_{x}}(\xi x)
H_{m_{y}}(\xi y)\right]^{2}
\eeq

Because of the relation between harmonic oscillator wave functions 
and their Fourier transforms, we see that there is a close relation 
between \( f_{0} \) and \( \rho(0) \).  This is given by
\beq
\lab{frho}
f_{0}=4\pi\left(\frac{\hbar^{2}}{ma^{2}}\frac{1}{k_{B}T_{0}}
\right)^{2}\rho(0).
\eeq

We can find WKB approximations for \( f_{k} \) and \( \rho(\rbf) \) by 
using the Boltzmann distribution function for the number of particles 
of wave number {\bf k} at position \( \rbf \), given in 2D by
\beq
f_{k}^{WKB}(\rbf)=\frac{1}{A}\frac{1}
{\exp\beta\left[\frac{\hbar^{2} k^{2}}{2m}+U(\rbf)-\mu\right]-1}.
\eeq
If we integrate over \( \rbf \) and sum over \( \kbf \) we should get 
the number of particles in the system.  However, in 2D we actually get 
just the \( F_{2}(\al) \) term in \Eq({2Dint}).  This is \( N^{\'} \), 
the number of particles in excited states.  We need to treat the 
condensate separately.

If we integrate \( f_{k}^{WKB}(\rbf) \) just over \( \rbf \) we get 
\( 
f_{k}^{WKB} \), the wave-number distribution, and if we just sum over 
\( \kbf \) we get \( \rho(\rbf) \), the local density.  The results of 
these integrations are
\beq
\lab{fWKB}
f_{k}^{WKB}=-\frac{2\hbar^{2}k_{B}T}
{ma^{2}(k_{B}T_{0})^{2}}\ln\left(1-\exp[-\beta
(\hbar k)^{2}/2m-\al]\right)
\eeq
and
\beq
\lab{rhoWKB}
\rho(\rbf)=-\frac{2\pi mk_{B}T}{h^{2}}
\ln\left(1-\exp[-\beta U(\rbf)-\al]\right)
\eeq
where \( \al = -\beta\mu \).

These formulas are valid only above any condensation or 
pseudo-critical temperature.  We could add \( \delta \)-function terms 
to account for the \( n_{0} \) term, but we do not bother here.  Were 
we to take $\kbf=0$ and consider low temperatures such that \( 
\al\propto 1/N \), the first of these functions would depend on 
$-\ln\al\propto \ln N$.  A similar behavior would occur for the second 
function for $\rbf=0$.  We will see next, however, that the true 
quantum behavior gives a stronger divergence with $N$.

In Fig.~4 we show plots of the quantum $f_k$, as a function of $k$ as 
computed from \Eq({fsubk}).  The lower temperature $0.5 T_0$ is below 
the transition temperature \( T_{c}=0.78 T_{0} \) and the higher 
temperature is above \( T_{c} \).  Also shown are the WKB 
approximations at the same temperatures from \Eq({fWKB}) and 
\Eq({rhoWKB}).  We see that the WKB approximation is quite good if we 
are above \( T_c \), but very poor
below as expected.

One important feature of $f_k$ shown is that $f_0$ is continuous with 
$f_k$ for $k\neq 0 $.  This is quite different from homogeneous Bose 
condensed systems for which the $k=0$ term is much larger that the 
$k\neq 0$ terms.  The pseudo-condensate in momentum space (terms not 
included in the WKB expression) {\em is spread over many $k$-values}. 
However, note that the \( k \)-scale factor determining the width of 
the spread in \( k\)-space of the functions in the Fig.~4 is 
\beq
\xi= \sqrt{\frac{mk_{B}T_{0}}{\hbar^{2}}}\frac{1}{N^{1/4}}
\eeq
which goes to zero as \( N\rightarrow\infty \) in the thermodynamic 
limit.

In Fig.~5, we show the $k=0$ momentum occupation number as a function 
of temperature for $N = 1000$.  Clearly there is a pseudo-condensation 
at the same temperature $T_c= 0.78T_0$ as the spatial condensation 
into the oscillator ground state.

The $N$-dependence of $f_0$ is shown in Fig.~6.  For the two lowest 
temperatures the behavior is found numerically to be 
$f_0\propto\sqrt{ 
N}$.  The highest temperature shown is just below the transition 
temperature and has not yet settled into its asymptotic form.  With 
this result we will have
\beq
\frac{f_{0}}{N}\propto\frac{1}{\sqrt{N}}\rightarrow 0
\eeq
as $N\rightarrow \infty $.  This supports our claim that the $k=0$ 
condensate fractions vanishes in the thermodynamic limit.

We can verify this numerical result analytically.  The first term in 
the sum in \Eq({fzero}) is just $\sigma_{00}= n_0$ so that the first 
term is of order $\sqrt{N}$ when $n_0$ is of order $N$.  The other 
terms in the series are positive so that the whole sum is at least of 
order $\sqrt{N}$.  The quantity in square brackets in \Eq({fzero}) is 
always less than 1.  This implies that
\beq
f_{0}< 
\frac{4\hbar^{2}}{ma^{2}k_{B}T_{0}}\frac{1}
{\sqrt{N}}\sum_{m_{x},m_{y}}\sigma_{m_{x}m_{y}}
\eeq
Since the sum on the right is just \( N \), we have 
\beq
f_{0}< 
\frac{4\hbar^{2}}{ma^{2}k_{B}T_{0}}\sqrt{N}
\eeq
This result implies that \( f_{0}/N \) approaches zero with \( N \) as 
$1/\sqrt{N}$ as our numerical treatment showed.  Actually we can do 
somewhat better in our analysis.  The quantity in square brackets in 
\Eq({fzero}) can be shown (via the Sterling's approximation) to obey
\beq
\lab{Stir}
\left[\frac{m_{x}!m_{y}!}{2^{m_{x}}2^{m_{x}}
[(m_{x}/2)!(m_{y}/2)!]^{2}}\right]\approx \frac{2}{\pi 
\sqrt{m_{x}m_{y}}}
\eeq
for large \(m_{x}\) and \( m_{y} \).  If we separate off the $m_x, 
m_y 
= 0$ term, use \Eq({Stir}) for {\em all} \( m_{x},m_{y} \), and 
then replace the resulting sum by an integral, we find
\beq
\lab{fappr}
f_{0}\approx 
\frac{4\hbar^{2}}{ma^{2}k_{B}T_{0}}\frac{n_{0}}{\sqrt{N}} + 
f_{0}^{WKB}
\eeq
where \( f_{0}^{WKB} \) is gotten from \Eq({fWKB}).  The WKB term in 
\Eq({fappr}) diverges as $\ln N$.  In this approximation the 
$\sqrt{N}$ behavior comes from just the term in $n_0/\sqrt{N}$.  We 
see from \Eq({fappr}), that for low \( T \), \( f_{0}\propto 
\sqrt{N}/a^{2}T_{0}\propto \sqrt{N\rho} \).  Thus \( f_{0}/N \) goes 
to zero in the thermodynamic limit as \( 1/\sqrt{N} \).  However, when 
the potential remains fixed, so that the density increases with \( N 
\), then \( f_{0}\propto N \).

Because of the relation \Eq({frho}), we see that the density at the 
origin also diverges as $\sqrt{N}$ in the thermodynamic limit.

It is possible to extend our discussion to arbitrary dimension.  We 
find that in any dimension \( f_{0}/N \) tends to zero as \( 
1/\sqrt{N} \) so that in the \ther there is no BEC into the \( k=0 \) 
state.  While the Hohenberg theorem forbids a $k=0$ condensate only 
for for D\( \leq 2 \), there is, on the other hand, there is no 
theorem that {\em requires} a BEC into \( k=0 \) for D\( >2\).

We should again point out that our results apply only to the ideal 
gas.  For a real gas with hard-core interactions there can be no 
divergence in the density.  If the particles have a hard-core 
interaction, the density will stay finite everywhere and the 
connection between the CPO and Hohenberg theorem will be 
re-established.  This implies that the 2D BEC into the lowest 
oscillator state will disappear for interacting particles in the 
thermodynamic limit.  Thus the CPO theorem implies that for 
interacting finite 2D systems the dependence of the transition 
temperature on particle number must change from \( T_{c}^{(2)}\propto
\sqrt{\rho}\propto \sqrt{N} \) as shown for an ideal gas to 
something more weakly dependent on \( N \).  An example would be the 
form \( T_{c}\propto \sqrt{N}/\ln N \) analogous to the 1D gas, but 
the determination of the exact form, of course, requires explicit 
hard-core calculations. Curiously, the 3D, and even the 1D, 
dependences are not required to change. 

\vfill\eject
\vspace{24pt}
\vskip 12pt

\centerline{\bf V.~~Discussion}
\vskip 12pt

We have examined several features of BEC in harmonic potentials for 
the ideal gas.  We have shown that, in the thermodynamic limit, there 
is no condensation into the zero-momentum state in any number of 
dimensions.  This result occurs in spite of the fact that there is a 
BEC into the lowest oscillator state for dimension D$\geq 2$.

For 2D there is a peculiarity in finding a condensate in the lowest 
oscillator state in the thermodynamic limit given the usual 
connection, via the CPO theorem, between the $k = 0$ condensation and 
the condensation into any single-particle state.  The fact that the 
Hohenberg theorem requires that there be no real condensation via a 
phase transition into the $k = 0$ state in 2D would normally rule out 
the possibility of BEC into any other state.  However, we have seen 
that the density diverges at the origin in the oscillator case (in the 
thermodynamic limit), which invalidates the CPO proof.  Thus while 
there is no BEC into the $k = 0$ momentum state in 2D in the 
thermodymanic limit, there {\em is} one into the \( m=0 \) oscillator 
state.

The thermodynamic limit is an approximation to reality that might seem 
less reasonable in a harmonic trap than in homogeneous systems.  The 
numbers of particles in the initial real experiments were quite small 
-- only a few thousand to a few million.  Further, the mode by which 
one must take the thermodynamic limit, weakening the potential while 
increasing the particle number in such a way that the average density 
remains constant, might seem qualitatively different from enlarging 
the size of a box of particles at constant density.  In the box case 
it would seem that the enlargement is merely a change in the boundary 
conditions, which affects the minority of particles at the walls.  On 
the other hand, varying the size of the experimental harmonic trap 
affects every particle.  However, looked in another way we can say 
that all that the particles in either case know is the energy levels 
and their spacing.  In both the box and the harmonic trap the 
thermodynamic limit treats the transition to the case where the 
spacing becomes infinitesimally small.  The process is thus completely 
analogous in the two cases, simply with differing densities of states.

The experimental situation, whether it be with a box or a harmonic 
magnetic trap, usually has a fixed potential (which of course can be 
altered) with possibly varying numbers of particles; obviously the 
\ther is never approached.  Thus it becomes more useful to look at 
finite systems and how their properties depend on the number of 
particles $N$.  Although finite systems do not have actual phase 
transitions, they have pseudo-condensations in which a substantial 
fraction of the $N$ particles fall into the lowest oscillator state 
below some temperature whose value depends on $N$.  Real experiments 
involve such pseudo-transitions and we have shown that they can occur 
in any number of dimensions, even in 1D. Experimentalists using 
magnetic traps may have the capability of manipulating the traps to 
simulate 2D or even 1D.$^{\ref{Ket}}$

The finiteness of the systems in real experiments may be precisely 
what makes our results relevant.  Because there are so few particles 
in the real systems, they do indeed behave much like ideal gases.  Our 
finite-size ideal gas model has a density that is well-behaved at the 
origin without the unrealistic divergent behavior shown in the 
thermodynamic limit.  In our finite-sized model, we can investigate 
how such systems behave as \( N\) varies.  There are 
pseudo-condensations into both the lowest oscillator state and into 
the \( k=0 \) state, and the relevant variables have predicted 
dependences on \( N \). The 1D case is interesting because of the 
behavior of the transition temperature; the case of 2D is particularly 
interesting from the point of view of the CPO theorem.  It is possible 
that these \( N \) dependences could be measured experimentally and 
the predictions demonstrated.  What needs to be done in further research 
is to investigate the effect of hard-cores to check that, in the low 
density limit, the results we find persist. 

The CPO theorem can predict that the persistence of ideal gas behavior 
cannot be complete.  If the particles have a hard-core interaction, 
the density will stay finite everywhere even in the thermodynamic 
limit and the connection between the CPO and Hohenberg theorem will be 
re-established.  This implies that the 2D BEC into the lowest 
oscillator state will disappear for interacting particles in the 
thermodynamic limit.  The theorem thus implies that for interacting 
finite 2D systems the dependence of \( T_{c}^{(2)} \) on particle 
number must change from that shown by an ideal gas in such a way that 
it can vanish in the thermodynamic limit.

We note that is is possible that the prediction for \( f_{0} \) could 
be checked rather directly in experiments using a common method of 
measuring the condensate.  The magnetic trap is removed and the 
particle cloud expands.  Its ``shadow'' is then examined after it has 
blown up to a size large enough to be easily visible.  The spatial 
distribution after removal of the trap is determined by the velocity 
distribution.  The particles staying at the center of the cloud are 
those with zero momentum.
  
\vskip 12pt
\centerline{\bf Appendix}
\vskip 12pt

\Eqs({2Dint}) and (\ref{3Danaly}) give corrections to the large-\( N 
\) limit of the relation giving \( n_{0}(T,N) \).  It is useful to 
consider a more rigorous derivation than we have given above and to 
make connection with previous work on the subject.  In particular, 
\Eq({3Danaly}) looks different from the equivalent formula of 
Ref.~\ref{Grob} [Eq.~(4) in the first paper of the reference], in 
having for example a\( \frac{5}{2} \) in the \( F_{2} \) term instead 
of a \( \frac{3}{2} \).  We will see however that our formula agrees 
completely with the results of Ref.~\ref{Grob}.

We use the Euler-Maclaurin summation formula\( ^{\ref{Abram}} \) to change 
our sums to integrals.  There is more than one form of this formula.  
Eq.~(3.6.28) of Ref.~{\ref{Abram}} gives 
\beqa 
\label{EM1}
\sum_{k=1}^{n-1}f(k)&=&\int_{0}^{n}f(k)dk-\frac{1}{2}[f(0)+f(n)]+ 
\frac{1}{12}[f'(n)-f'(0)]\nonumber\\ 
&-& \frac{1}{720}[f'''(n)-f'''(0)]+\cdots 
\eeqa
Note that the sum starts at \( k=1 \). \Eq({N3D}) can be written as 
\beq
N=n_{0}+\sum_{p=1}^{\infty}\frac{\frac{1}{2}p^{2}+
\frac{3}{2}p+1}{e^{bp+\al'}-1}
\eeq
where 
\beq
b=\frac{T_{0}^{(3)}}{TN^{1/3}}.
\eeq
Use of \Eq({EM1}) leads to 
\beq
N=n_0 + \frac{1}{b^{3}}F_{3}(\al)+\frac{3}{2}\frac{1}{b^{2}}F_{2}(\al)
+\cdots
\eeq
exactly as in Ref.~\ref{Grob}.  However, the corrections to the 
Euler-Maclaurin integral form a divergent series.  It is easy to see 
that the various derivative terms are in powers of \( 
n_{0}=1/[\exp (\al)-1]\), which for low \( T \) is of order \( N \).

To find a correction series that converges we use a slightly different 
form of the Euler-Maclaurin formula\( ^{\ref{Haug}} \); Eq.~(23.1.30) 
of Ref.~\ref{Abram} is
\beqa
\label{EM2}
\sum_{k=0}^{n}f(k)&=&\int_{0}^{n}f(k)dk+\frac{1}{2}[f(0)+f(n)]+ 
\frac{1}{12}[f'(n)-f'(0)]\nonumber\\ 
&-& \frac{1}{720}[f'''(n)-f'''(0)]+\cdots 
\eeqa 
This form, in which the sum starts at \( k=0 \), can be used directly on 
\Eq({Nseries}). Using it we get 
\beqa
N&=&n_0 +\frac{1}{b^{3}}F_{3}(\al')
+\frac{5}{2}\frac{1}{b^{2}}F_{2}(\al')+3 \frac{1}{b}F_{1}(\al')
\nonumber\\
&+&\frac{3}{2}\frac{1}{e^{\al'}-1}-\frac{1}{12}
\left[\frac{5/2}{e^{\al'}-1}-\frac{3}{(e^{\al'}-1)^{2}}\right]+\cdots 
\eeqa
Because \( \al'=\al+b \) appears in the arguments instead of \( \al \) 
the series of corrections is well-behaved as we now see.

First, we consider the case when \( T>T_{c} \) so that \( \al \) is of 
order unity while \( b \) remains of order \( N^{-1/3} \). We can 
simply Taylor expand around \( \al'=\al \). We make use of the easily 
proved relation
\beq
\frac{d}{d\al}F_{\sigma}(\al) = -F_{\sigma-1}(\al)
\eeq
for \( \sigma >1 \). After a bit of algebra we find to the same order
\beqa
N&=&n_0 +\frac{1}{b^{3}}F_{3}(\al)
+\frac{3}{2}\frac{1}{b^{2}}F_{2}(\al)+\frac{1}{b}F_{1}(\al)
\nonumber\\
&=& +\frac{31}{24}n_{0}+\frac{1}{12}n_{0}^{2}
\left[3-\frac{31}{2}b\right]-\frac{1}{2}bn_{0}^{3}+\cdots 
\eeqa
Since we are above \( T_{c} \), \( n_{0} \) is of order unity; \( b 
\) is of order \( N^{-1/3} \) so that the entire series in various 
powers of \( n_{0} \) and \( b \) is negligible. We have then
\beq
N\approx \frac{1}{b^{3}}F_{3}(\al)
+\frac{3}{2}\frac{1}{b^{2}}F_{2}(\al)+\frac{1}{b}F_{1}(\al)
\eeq
in agreement with Refs.~\ref{Grob}. Ref.~\ref{Haug} has given a 
similar analysis. 

Finally, below \( T_{c} \), \( \al \) is of order \( 1/N \), which is 
much smaller than \( b \), so that \( \al'\approx b \) and 
\beq
\frac{1}{e^{\al'}-1}\approx 1/b = O(N^{1/3}).
\eeq
The full Taylor expansion then gives 
\beq
N=n_{0}+N \frac{1}{b^{3}}F_{3}(0)
+\frac{3}{2}\frac{1}{b^{2}}F_{2}(0)+O\left(\frac{1}{b}\right)
\eeq 
The first three terms correspond to the formula of Ref.~\ref{Grob}, 
which is therefore valid below the transition as well as above.

\vskip 12pt
\centerline{\bf Acknowledgement}
\vskip 12pt
We wish to thank Dr. F. Lalo\( \ddot{\rm e} \) for useful conversations.

\vfill\eject
\centerline{\bf References}
\begin{enumerate}

\item\label{Corn}M. H.  Anderson, J. R.  Ensher, M. R.  Matthews, C. 
E. Wieman, and E. A. Cornell, Science {\bf 269}, 198 (1995).

\item\label{Ketb}K. B.  Davis, M.-O.  Mewes, M. R.  Andrews, N. J.  
van Druten, D. S. Durfee, D. M. Kurn, and W. Ketterle, Phys.  Rev.  
{\bf 75}, 3969 (1995).

\item\label{Bra}C.  C.  Bradley, C.  A.  Sachett, J. J.  Tollett, and 
R. G.  Hulet, Phys.  Rev.  {\bf 75}, 1687 (1995).

\item\label{Osb}M. F. M.  Osborne, Phys.  Rev.  {\bf 76}, 396 (1949).

\item\label{Zim}J. M.  Ziman, Phil.  Mag.  {\bf 44}, 548 (1953).

\item\label{Mil}D. L. Mills, Phys. Rev. {\bf 134}, A306 (1964).

\item\label{Gob}D. F.  Goble and L. E.  H.  Trainor, Can. J. Phys. 
{\bf 44}, 27 (1966); Phys.  Rev.  {\bf 157}, 167 (1967).

\item\label{Kru}D. A. Kruger, Phys. Rev. {\bf 172}, 211 (1968).

\item\label{Imr}Y. Imry, Ann. Phys., {\bf 51}, 1 (1969).

\item\label{Bar}M. N.  Barber, in {\it Phase Transitions and Critical 
Phenomena}, ed.  C.  Domb and M. S.  Green (Academic Press, London, 
1983), Vol.  8, p.  146.

\item\label{Gro}S. Grossmann and M. Holthaus, Z. Naturforsch. {\bf 
50a}, 323 (1995); Z. Phys. B {\bf 97}, 319 (1995).

\item\label{Mul} W. J. Mullin, ``Pseudo-Bose Condensation in Finite 
Ideal Systems'' unpublished report, 1981.

\item\label{Ket}W.  Ketterle and N. J.  van Druten, Phys.  Rev. A (to 
be published).

\item\label{Hoh}P.  C.  Hohenberg, Phys.  Rev.  {\bf 158}, 383 (1967).

\item\label{Chesb}C. V.  Chester, in {\it Lectures in Theoretical 
Physics } ed.  K. T.  Mahanthappa (Gordon and Breach, Science 
Publishers, Inc., New York, 1968) Vol.  11B, p.  253

\item\label{Chesa}C. V.  Chester, M. E.  Fisher, and N. D.  Mermin, 
Phys.  Rev.  {\bf 185}, 760 (1969).

\item\label{Pen}O. Penrose and L.  Onsager, , Phys.  Rev.  {\bf 104}, 
576 (1956).

\item\label{Reh}J. J.  Rehr and N. D.  Mermin, Phys.  Rev.  B {\bf 
185}, 3160 (1970).

\item\label{Mas}R.  Masut and W. J.  Mullin, Am.  J.  Phys. {\bf 47}, 
493 (1979).

\item\label{deG}S.  deGroot, G. J.  Hooyman, and C. A.  ten Seldam, 
Proc.  R.  Soc.  (London) A, {\bf 203}, 266 (1950).

\item\label{SDS}C. E.  Campbell, J. G.  Dash, and M.  Schick, Phys.  
Rev.  {\bf 26}, 966 (1971).

\item\label{Klepa}V.  Bagnato, D.  Pritchard, and D.  Kleppner, 
Phys.  Rev.  A {\bf 35}, 4354 (1987).

\item\label{Klepb}V.  Bagnato and D.  Kleppner, Phys.  Rev.  
A {\bf 44}, 7439 (1991).  

\item\label{Grob}S. Grossmann and M. Holthaus, Phys Lett. A {\bf 208}, 
188 (1995); Z. Naturforsch. {\bf 50a}, 921 (1995).

\item\label{Krauth}W. Krauth, (preprint).

\item\label{Tcexper} M.-O. Mewes, M. R. Andrews, N. J. van Druten, D. 
M. Kurn, D. S. Durfee, and W. Ketterle, preprint.
  
\item\label{Haug}H. Haugerud and F. Ravndal, cond-mat/9509041. 

\item\label{Haugb}H. Haugerud, T. Haugset, and F. Ravndal, 
cond-mat/9605100.

\item\label{Kirst} K. Kirsten and D. J. Toms, Phys. Lett. B {\b 368}, 
119 (1996); cond-mat/9604031; cond-mat/9607047; cond-mat/9608032.

\item\label{Wid}A. Widom, Phys.  Rev. {\bf 168}, 150 (1968).  

\item\label{Lon}F. London, {\it Superfluids} (Dover, New York, 1964) 
Vol. II, p. 203; J. E. Robinson, Phys. Rev. {\bf 83}, 678 (1951).

\item\label{Abram}M. Abramowitz and I. A. Stegun {\it Handbook of 
Mathematical Functions} (Dover, New York, 1972).     

\end{enumerate}

\vfill\eject
\vskip 12pt

\centerline{\bf Figure Captions}

\vskip 12pt
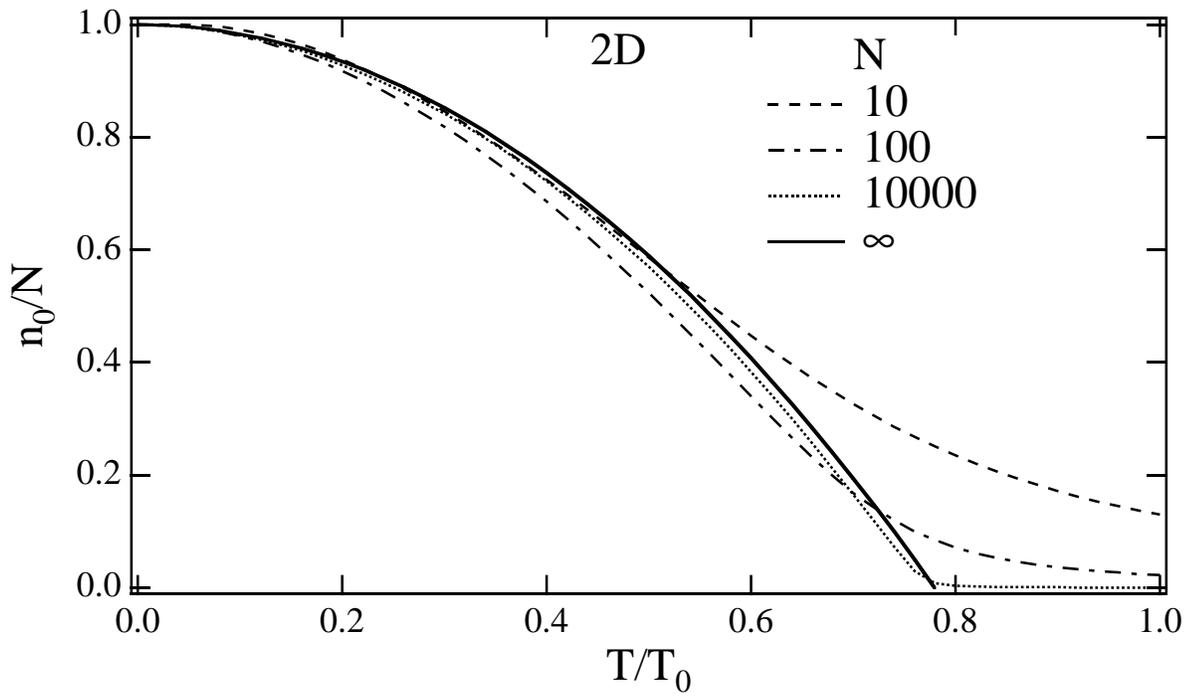
\noindent Figure 1: Spatial condensate fraction $n_0/N$ versus 
temperature ($T/T_0$) in 2D.  The dotted and dashed lines are for 
finite $N$ values, generated from \Eq({2Dpeq}); the solid curve is the 
thermodynamic limit, \Eq({n02D}). \( T_{0} \) is given by \Eq({T0}).\\

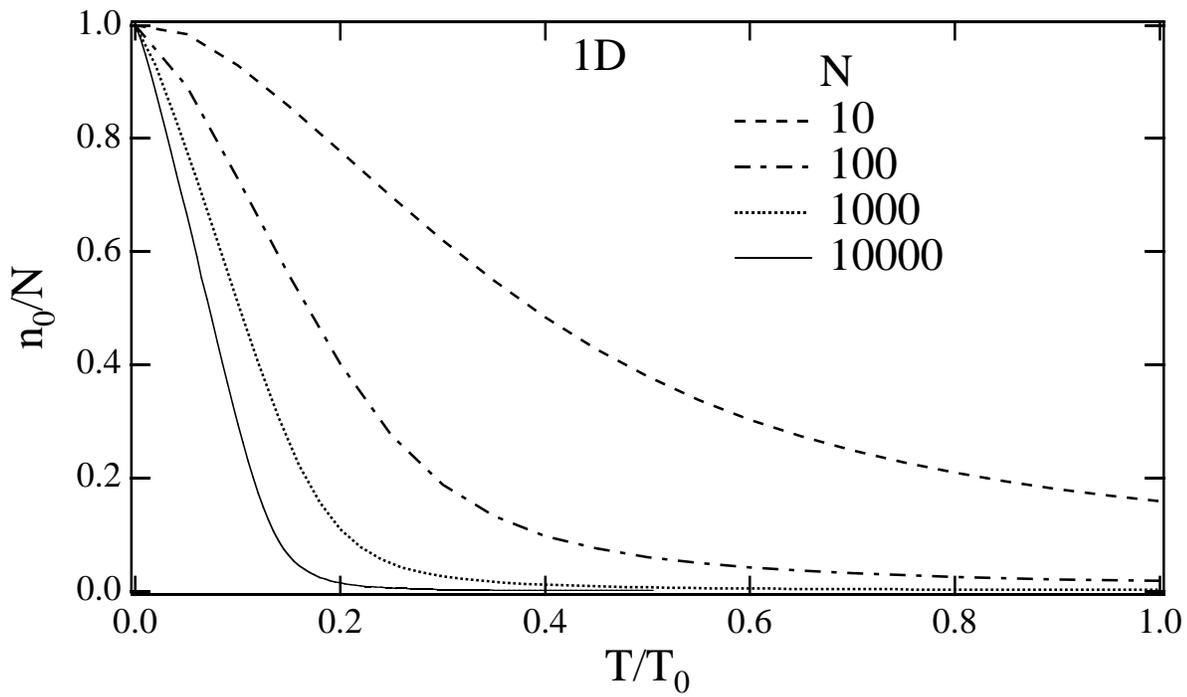
\noindent Figure 2: Spatial condensate fraction $n_0/N$ versus $T/T_0$ 
in 1D from \Eq({1Dsum}).  This curve is appropriate for considering 
the thermodynamic limit for which the average density and thus $T_0$ 
are held constant by weakening the potential as $N $ increases.  The 
pseudo-critical temperature then decreases with $N$ as $1/\ln N$. 
In 1D \( T_{0} \) is given by \Eq({T0D}) with D=1.\\

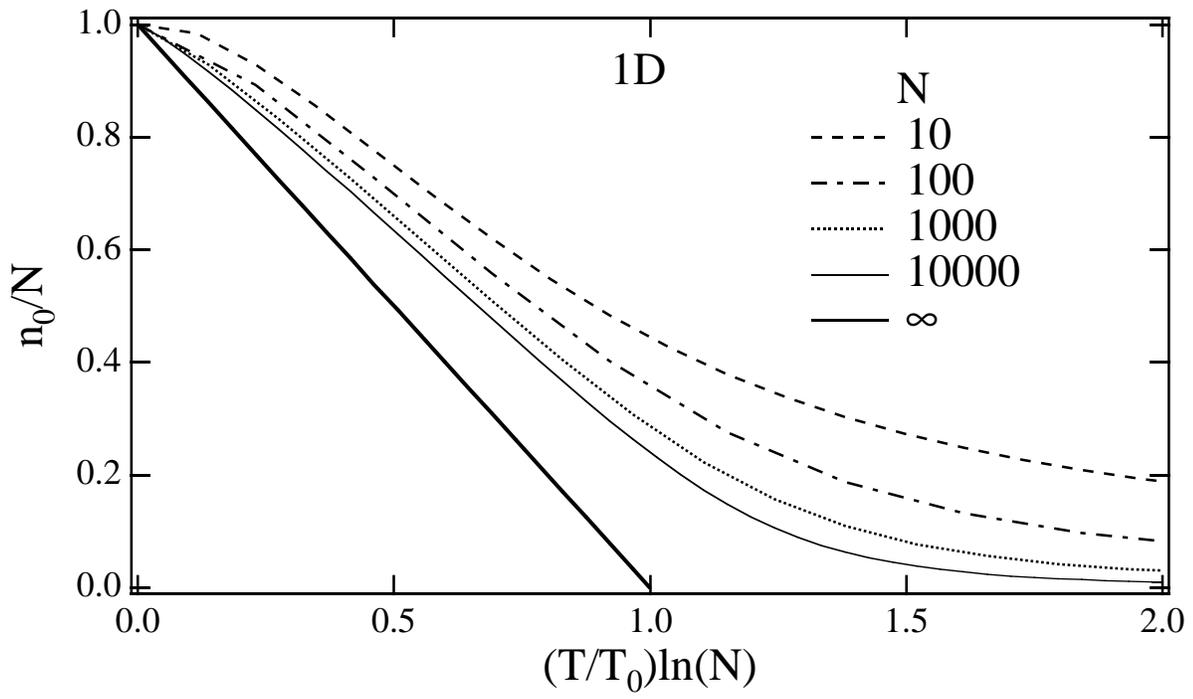
\noindent Figure 3: Spatial condensate fraction $n_0/N$ versus 
$(T/T_0)\ln N$ in 1D from \Eq({1Dsum}).  This plot removes the $N$ 
dependence from the approximate linear expression, \Eq({1Dn0}).  As 
$N$ increases the exact expressions approach the linear limit.  When 
the harmonic potential is fixed while $N$ increases, as in real 
experiments, $T_0/\ln N\propto N/\ln N$; that is, the pseudo-critical 
temperature increases with $N$.\\

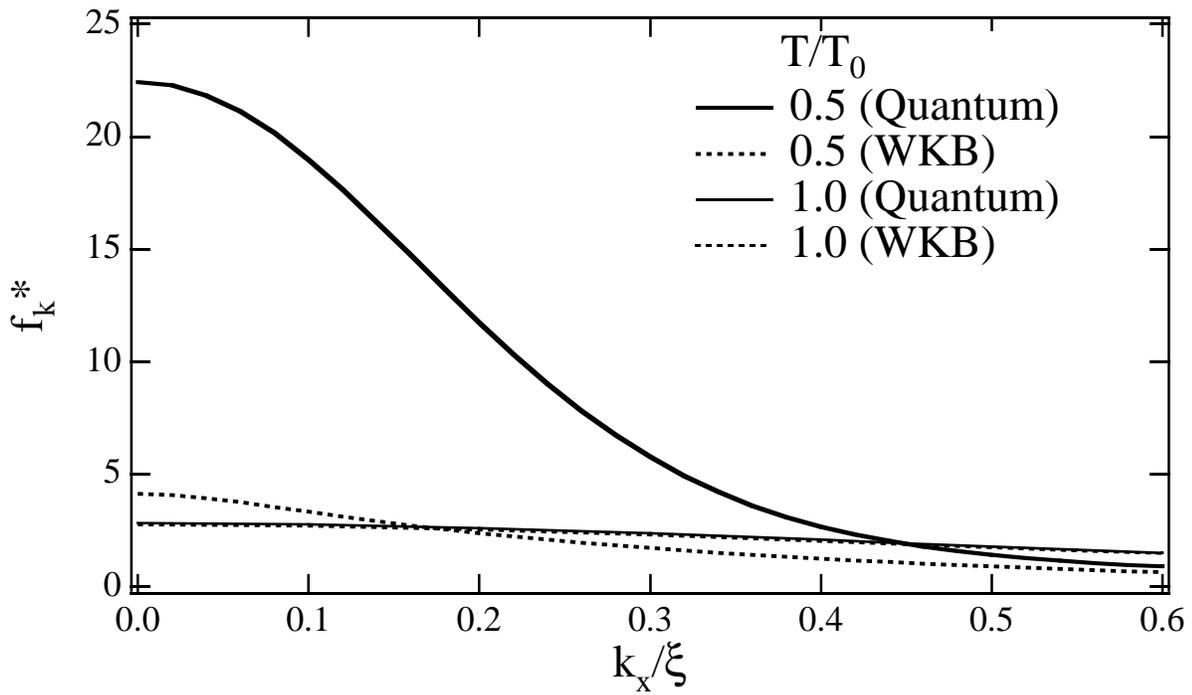
\noindent Figure 4: Reduced momentum distribution function 
$f_k^{\ast}$ versus reduced momentum component $k_x/\xi$ in 2D.  Here 
$f_k^{\ast}\equiv f_k[ma^2k_BT_0/\hbar^2]$ and we have taken 
$k_y=k_x$.  The upper solid curve is the exact quantum results from 
\Eq({fsubk}) for a temperature above $T_c$ and the lower one for a 
temperature below $T_c$.  The dotted curves are the WKB approximations 
at the same temperatures as the solid curves.  The WKB expression is a 
good approximation only if $T>T_c$.\\

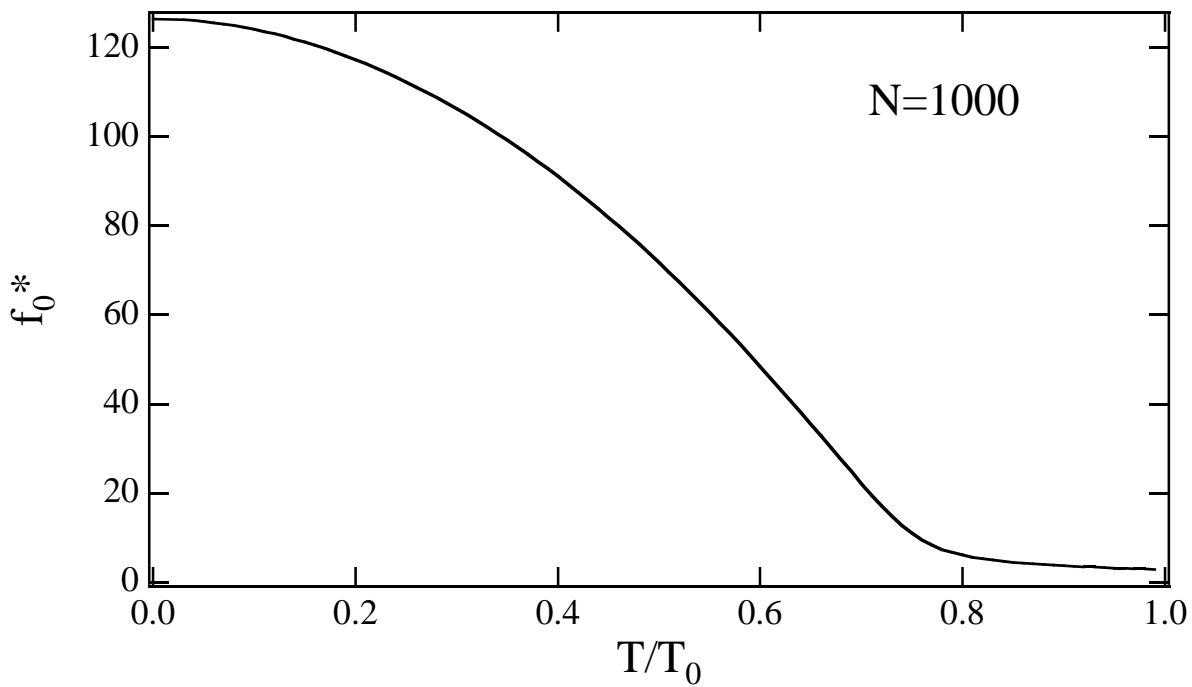
\noindent Figure 5: Reduced zero-momentum distribution $f_0^{\ast}$ 
versus $T/T_0$ in 2D from \Eq({fzero}) in 2D for $N=1000$.  Here 
$f_k^{\ast}\equiv f_k[ma^2k_BT_0/\hbar^2]$.  A pseudo-condensate in 
momentum space clearly occurs for $T<T_c \approx 0.78T_0$.\\

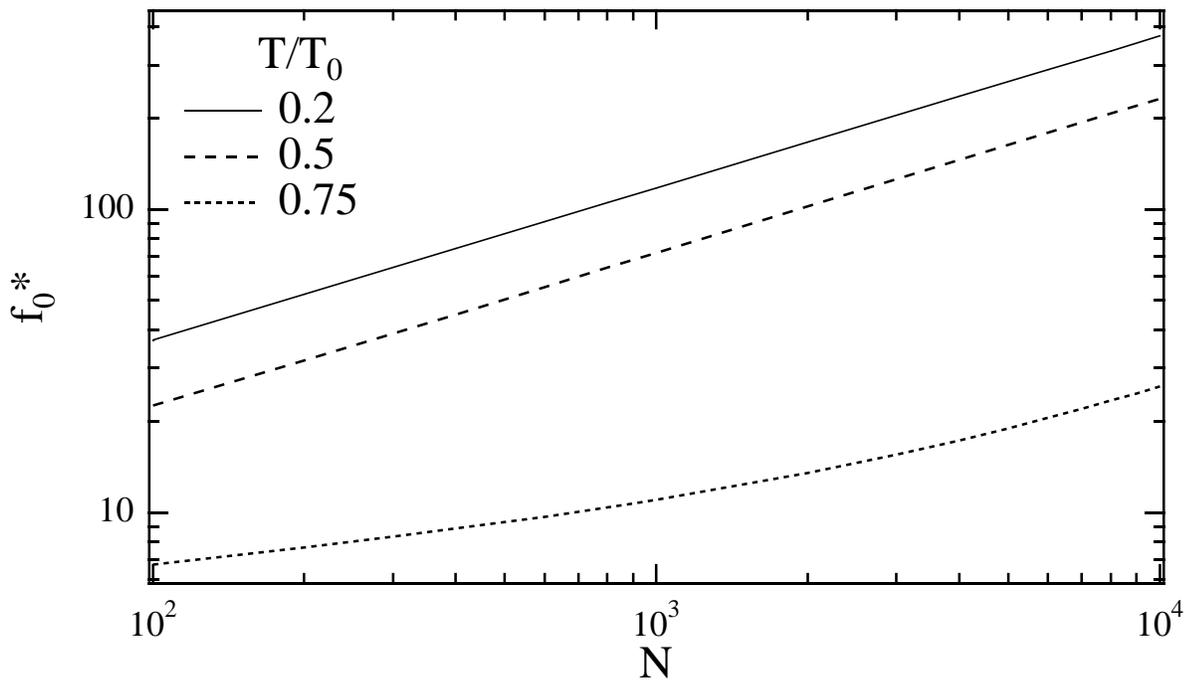
\noindent Figure 6: Reduced zero-momentum distribution $f_0^{\ast}$ 
versus $N$ in 2D [\Eq({fzero})] for several temperatures below $T_c$.  
These are log-log plots, with slopes of the upper two curves 
numerically very close to 0.5, indicating that $f_0\propto \sqrt{N}$.  
The lowest curve, for $T$ just below $T_c$, has not yet reached its 
asymptotic $N$ dependence.

\end{document}